\documentclass[twocolumn,showpacs,superscriptaddress,amssymb,10pt,pra,floatfix]{revtex4}

\usepackage{graphicx}
\usepackage{dcolumn}
\usepackage{bm}
\usepackage{epsfig}
\usepackage{color}
\usepackage{longtable}
\date{December 24, 2016}

\usepackage[cp866]{inputenc}
 \usepackage[english]{babel}
 \newcommand{\be}{\begin{equation}}
\newcommand{\ee}{\end{equation}}
\newcommand{\dst}{\displaystyle}
\newcommand{\fr}[2]{\frac{{\dst #1}}{{\dst #2}}}

\newcommand{\bea}{\begin{eqnarray}}
\newcommand{\eea}{\end{eqnarray}}
\newcommand{\nn}{\nonumber}
\def\sprm#1#2{  \left\langle #1 \left\vert \right. #2 \right\rangle   }

\begin{document}

\preprint{}
%
%
%
%
\title{Scattering of twisted electron wave--packets
by atoms in the Born approximation}
%
%
%
%

\author{D.~V.~Karlovets}
\affiliation{Department of Physics, Tomsk State University, Lenina Ave. 36, 634050 Tomsk, Russia}

\author{G.~L.~Kotkin}
\affiliation{Novosibirsk State University, RUS--630090, Novosibirsk, Russia}
\affiliation{Sobolev Institute of Mathematics, RUS-630090, Novosibirsk, Russia}

\author{V.~G.~Serbo}
\affiliation{Novosibirsk State University, RUS--630090, Novosibirsk, Russia}
\affiliation{Sobolev Institute of Mathematics, RUS-630090, Novosibirsk, Russia}

\author{A.~Surzhykov}
\affiliation{Physikalisch--Technische Bundesanstalt, D--38116 Braunschweig, Germany}
\affiliation{Technische Universit\"at Braunschweig, D--38106 Braunschweig, Germany}

\date{\today \\[0.3cm]}

%
%
%
%
%
\begin{abstract}
The potential scattering of electrons carrying non--zero quanta of the orbital angular momentum (OAM) is studied in a framework of the generalized Born approximation, developed in our recent paper
by Karlovets \textit{et al.}, Phys. Rev. A. {\textbf 92}, 052703 (2015). We treat these so--called \textit{twisted} electrons as spatially localized wave--packets. The simple and convenient expressions are derived for a number of scattering events in collision of such a vortex electron with a single potential, located at a given impact parameter with respect to the wave-packet's axis. The more realistic scenarios are also considered with either localized (mesoscopic) targets or infinitely wide (macroscopic) ones that consist of the randomly distributed atoms. Dependence of the electron scattering pattern on a size and on a relative position of the target is studied in detail for all three scenarios of the single--potential--, mesoscopic-- and the macroscopic targets made of hydrogen in the ground $1s$ state. The results demonstrate that the angular distribution of the outgoing electrons can be very sensitive to the OAM and to kinematic parameters of the focused twisted beams, as well as to composition of the target. Scattering of vortex electrons by atoms
can, therefore, serve as a valuable tool for diagnostic of such beams.
\end{abstract}

\pacs{34.10.+x, 34.80.-i, 42.50.Tx}
\maketitle

%
%
\section{Introduction}

Beams of electrons carrying non--zero projection of the orbital angular momentum (OAM) upon their propagation direction currently attract much interest, both in theory and in experiment \cite{BlB07,UcT10,VeT10}. A wave--front of these electrons has a helical spatial structure which twists around the beam axis. Such \textit{twisted} beams are produced today at scanning electron microscopes with energies up to a few hundreds of keV and with the OAM projection as high as $\hbar m = 1000 \hbar$ \cite{VeT10,GrG15,MaG15,m=1000}. Owing to the non-vanishing $m$, the twisted electrons possess a magnetic moment $\mu \propto m \mu_B$ with $\mu_B$ being the Bohr magneton. This -- additional (and big when $m \gg 1$)-- magnetic moment makes the OAM beams particularly suitable for probing magnetic properties of materials at the nano-- and even atomic--scale \cite{VeT10,RuB14,ErL16} as well as for studying unusual properties of the electromagnetic radiation generated by such electrons \cite{Iv13}.

A number of material studies are planned with the twisted beams that are based on the (analysis of) electron \textit{scattering} by targets. In order to achieve high spatial resolution in these scattering experiments one would need to focus electron beams to a sub--nanometer scale \cite{ErL16}. Such tight focusing implies that the relative \textit{size} and \textit{position} of the beam and a target can be of paramount importance. Detailed theoretical analysis of the beam--size (and position) effects is demanded, therefore, for the guidance and analysis of the twisted--electron scattering experiments. First steps towards such an analysis were carried out by one of us \cite{IvS11} as well as by Van Boxem and co--workers \cite{BoP14,BoP15}. Those previous works, however, mainly dealt with the scattering amplitudes and angular distributions of the outgoing electrons. Much less attention has been paid so far to the \textit{quantitative} study of the scattering process which requires evaluation of (i) the number of events and, if possible, (ii) the cross sections. The analysis of these (quantitative) observables is a rather complicated task which requires accurate account of the beam's (and the target's) sizes, as well as of the position effects.

In this contribution we derive and study the number of events and the
generalized cross sections for potential scattering of the focused twisted electron beams.
Such beams are treated as spatially localized wave--packets that collide with targets
consisting of short--range potentials. The theoretical analysis of these collisions is
based on the generalized Born approximation which was recently elaborated by us in Ref.~\cite{KaK16} and was successfully used for describing scattering of ``the ordinary'' Gaussian packets.

In order to apply the (generalized) Born theory for twisted electrons, we first construct their wave--packets in Section~\ref{subsec:twisted_states}. The scattering amplitude for a single potential placed at a particular impact parameter is derived then in Section~\ref{subsec:amplitudes}. With the help of this amplitude and by making use of the theory from Ref.~\cite{KaK16}, we obtain the number of events and the averaged angle-differential cross sections for different ``experimental set--ups''. For example, in Section ~\ref{subsec:cross_sections_number_of_events} we consider collisions of a twisted electron
with a single scatterer as well as with the localized- (mesoscopic-) and the infinitely wide (macroscopic) targets, the latter are described as incoherent ensembles of potential centers.
The derived expressions for the number of events and the averaged cross sections can be applied for a target consisting of central short--range potentials of \textit{any} form. In Section~\ref{sec:Yukawa_and_hydrogenic_formulas}, however, we discuss the particular cases of Yukawa and hydrogenic potentials. In collision theory, these potentials are often used to approximate realistic electron--atom interactions, see for example Refs.~\cite{SaM87,Sal91}. Detailed calculations for the Yukawa-- and hydrogenic targets and for fast but still non-relativistic electrons are presented in Section~\ref{sec:results}.
Throughout these calculations, emphasis is placed, in particular, on the questions of (i) whether the number of scattering events in collision of twisted electrons is comparable to the one in the ``standard'' plane--wave case, (ii) how the averaged cross section depends on a size of the initial wave packet, and (iii) how the angular distributions of outgoing electrons depend on the impact parameters $b$ between the potential centre and the packet's axis. The summary of these results and outlook will be given in Section~\ref{sec:summary}.

Similar to the previous work~\cite{KaK16}, we use units with the Planck's constant set to unity, $\hbar = 1$.

%
%
\section{Theoretical background}
\label{sec:theory}

%
%
\subsection{Twisted electron states}
\label{subsec:twisted_states}

Before we start with the analysis of the potential scattering of twisted electrons, let us briefly remind how these electrons are described within the non--relativistic framework. First we will discuss the monochromatic Bessel solutions, and will show later how to construct the (twisted) wave--packets.

\subsubsection{Monochromatic Bessel states}

Not much has to be said about the stationary electron states with the well--defined energy $\varepsilon$, longitudinal momentum $k_z$ and projection $m$ of the orbital angular momentum onto the quantization ($z$--) axis. In the cylindrical coordinates ${\bm r} = (r_\perp, \varphi_r, z)$, these \textit{Bessel} solutions of the field--free Schr\"odinger equation are described by the wave--function \cite{JeS11,JeS11a,IvS11}:
\begin{equation}
   \label{eq:wavefunction_Bessel_stationary}
   \sprm{{\bm r}}{k_z \, \varkappa \, m} = {\rm e}^{-i \omega t + i k_z z} \;
   \Psi_{\rm tr}^{\varkappa m}({\bm r}_\perp) \, ,
\end{equation}
with the transverse component
\begin{equation}
   \label{eq:wavefunction_Bessel_stationary_transverse}
   \Psi_{\rm tr}^{\varkappa m}({\bm r}_\perp) =
   \frac{{\rm e}^{i m \varphi_r}}{\sqrt{2\pi}} \;
   \sqrt{\varkappa}\, J_m(\varkappa\, r_\perp) \, ,
\end{equation}
where $J_m(\varkappa\, r_\perp)$ is the Bessel function and the absolute value of the transverse momentum is fixed to $|{\bm k}_\perp| = \varkappa = \sqrt{2m_e\,\varepsilon- k_z^2}$.

For practical calculations, it is often more convenient to write the function $\sprm{{\bm r}}{k_z \, \varkappa \, m}$ in the momentum representation:
\begin{equation}
   \label{eq:wavefunction_Bessel_stationary_momentum}
   \sprm{{\bm r}}{k_z \, \varkappa \, m} = {\rm e}^{-i \omega t + i k_z z} \, \int
   \frac{{\rm d}^2{\bm k}_\perp}{(2\pi)^2} \,
   \Phi_{\rm tr}^{\varkappa m}({\bm k}_\perp) \, {\rm e}^{i {\bm k}_\perp {\bm r}_\perp} \, .
\end{equation}
Here, the Fourier coefficient $\Phi_{\rm tr}^{\varkappa m}({\bm k}_\perp)$ is the transverse part of the momentum--space wave--function and is given by:
\begin{equation}
   \label{eq:wavefunction_Bessel_stationary_momentum_transverse}
   \Phi_{\rm tr}^{\varkappa m}({\bm k}_\perp) =
   (-i)^m \, \frac{{\rm e}^{i m \varphi_k}}{\sqrt{2\pi}}
   \,\frac{\delta(k_\perp - \varkappa)}{\sqrt{k_\perp}} \, .
\end{equation}
As can be seen from these two expressions, the Bessel electron state can be \textit{understood} as a coherent superposition of the plane waves, whose wave--vectors ${\bm k} = ({\bm k}_\perp, k_z)$ lay on the surface of a cone with the opening angle $\tan \theta_k = \varkappa / k_z$.

%
%
\subsubsection{Wave--packets of Bessel states}

The monochromatic Bessel state, (\ref{eq:wavefunction_Bessel_stationary})--(\ref{eq:wavefunction_Bessel_stationary_transverse}) and (\ref{eq:wavefunction_Bessel_stationary_momentum})--(\ref{eq:wavefunction_Bessel_stationary_momentum_transverse}), is a non--square--integrable solution for the Schr\"odinger equation that is spread over the \textit{entire} coordinate space, i.e., its dispersions are $\Delta z = \Delta {r_\perp} = \infty$. Such a monochromatic solution is not sufficient to describe nowadays experiments in which electron beams can be focused to a sub--nanometer scale. Therefore, we have to employ $\sprm{{\bm r}}{k_z \, \varkappa \, m}$ in order to construct a \textit{spatially--localized} wave--packet:
\begin{eqnarray}
   \label{eq:wavepacket_Bessel_coordinate}
   \Psi_{\varkappa_0 m p_i}({\bm r}) &=& {\rm e}^{-i \omega t} \, \Psi^{(m)}_{\rm tr}({\bm r}_\perp) \nonumber\\[0.1cm]
   &\times& \int_0^\infty {\rm e}^{i k_z z} \, g_{p_i \sigma_{k_z}}(k_z) \, {\rm d}k_z  \, ,
\end{eqnarray}
where the convolution of the transverse component (\ref{eq:wavefunction_Bessel_stationary_transverse}) reads as:
\begin{eqnarray}
   \label{eq:transverse_component_convolution}
   \Psi^{(m)}_{\rm tr}({\bm r}_\perp) &=&  \int_0^\infty \Psi_{\rm tr}^{\varkappa m}({\bm r}_\perp) \, g_{\varkappa_0 \sigma_\varkappa}(\varkappa) \, {\rm d}\varkappa \nonumber \\
   &=& \frac{{\rm e}^{i m \varphi_r}}{\sqrt{2\pi}} \,
   R^{(m)}(r_\perp) \, ,
\end{eqnarray}
with $g_{p_i \sigma_{k_z}}(k_z)$ and $g_{\varkappa_0 \sigma_\varkappa}(\varkappa)$ being the weight functions, and
\begin{equation}
\label{eq:R_function}
R^{(m)}\left(r_\perp\right) = \int_0^\infty
\sqrt{\varkappa} \, J_m\left(\varkappa\, r_\perp \right) \, g_{\varkappa_0 \sigma_\varkappa}(\varkappa) \, {\rm d}\varkappa \, .
\end{equation}
The wave--packet $\Psi_{\varkappa_0 m p_i}({\bm r})$ is a physically--relevant solution with the definite OAM's projection $m$, with the mean values of the longitudinal momentum
$\left< k_z \right> = p_i$ and of the absolute value of the transverse one, $\left< k_\perp \right> = \varkappa_0$.

As before, we can represent the transverse part of the wave--packet (\ref{eq:transverse_component_convolution}) in terms of the plane--wave components:
\begin{eqnarray}
   \label{eq:wavepacket_Bessel_momentum}
   \Psi_{\varkappa_0 m p_i}({\bm r}) &=&
   {\rm e}^{-i \omega t} \, \int \frac{{\rm d}^2{\bm k}_\perp}{(2\pi)^2} \,
   \Phi_{\rm tr}^{(m)}({\bm k}_\perp) \, {\rm e}^{i {\bm k}_\perp {\bm r}_\perp} \nonumber\\[0.1cm]
   &\times& \int_0^\infty {\rm e}^{i k_z z} \, g_{p_i \sigma_{k_z}}(k_z) \, {\rm d}k_z  \, ,
\end{eqnarray}
where the convoluted Fourier coefficient is
\begin{eqnarray}
   \label{eq:wave_packet_transverse}
   \Phi_{\rm tr}^{(m)}({\bm k}_\perp) &=&
   \int_0^\infty \Phi_{\rm tr}^{\varkappa m} ({\bm k}_\perp) g_{\varkappa_0 \sigma_\varkappa}(\varkappa) \, {\rm d}\varkappa \nonumber \\
   &=& (-i)^m\,\frac{e^{im\varphi_k}}{\sqrt{2\pi k_\perp}}\, g_{\varkappa_0 \sigma_\varkappa}(k_\perp) \, .
\end{eqnarray}
The weight functions $g_{p_i \sigma_{k_z}}(k_z)$ and $g_{\varkappa_0 \sigma_\varkappa}(\varkappa)$ in Eqs.~(\ref{eq:wavepacket_Bessel_coordinate})--(\ref{eq:wave_packet_transverse}) are peaked around $p_i$ and $\varkappa_0$ and have the widths $\sigma_{k_z}$ and $\sigma_\varkappa$, respectively. For the numerical analysis below we choose these functions to be of the Gaussian form.
For example, the distribution of the transverse momentum is described by:
\begin{equation}
   \label{eq:Gausian_distribution}
   g_{\varkappa_0 \sigma_\varkappa}(\varkappa) =
   C \, {\rm e}^{-(\varkappa - \varkappa_0)^2/(2\sigma^2_\varkappa)} \, ,
\end{equation}
where the constant $C$ is determined from the normalization condition $\int |g_{\varkappa_0 \sigma_\perp}(\varkappa)|^2 \, {\rm d}\varkappa = 1$. Eq.~(\ref{eq:Gausian_distribution}) corresponds to the dispersion $\Delta x = \Delta y \sim 1 /\sigma_{\varkappa}$ in the transverse plane. For the small values, $\sigma_\varkappa \ll 1/a$, where $a$ is the typical radius of the field action, the $\Delta x$ and $\Delta y$ become large and we call it a \textit{wide} wave--packet limit. The approximation of a wide packet will be employed below for simplifying the formulas for the scattering cross sections and the number of events.

%
%
\subsection{Scattering amplitudes}
\label{subsec:amplitudes}

Having briefly discussed construction of the Bessel electron's wave--packets, we are ready now to describe an amplitude for potential scattering. We consider the experimentally--relevant scenario in which (i) the longitudinal size of the packet $\sigma_z $ is larger than the characteristic radius of the field action, $\sigma_{z} \gg a$, and (ii) the dispersion of the packet in the transverse plane during the collision is negligible, that is, $t_{\rm dis} \sim \sigma_\perp/v_\perp \gg t_{\rm col} \sim \sigma_z/v_z$. Under these assumptions, which can be re--written as
\begin{equation}
   \label{eq:assumptions_for_amplitude}
   a\ll  \sigma_z \ll p_i/(\varkappa_0 \sigma_\varkappa) \, ,
\end{equation}
the scattering amplitude is given by Eqs.~(29)--(30) of Ref.~\cite{KaK16}:
\begin{equation}
   \label{eq:amplitude_monochromatic}
   F({\bm Q}, {\bm b}) = \int f({\bm Q} - {\bm k}_\perp) \,
   \Phi_{\rm tr}^{\rm (m)}({\bm k}_\perp) \, {\rm e}^{i {\bm k}_\perp {\bm b}} \, \frac{{\rm d}^2 {\bm k}_\perp}{2 \pi} \, .
\end{equation}
In this expression, the convoluted transverse component of the wave--packet is given by Eq.~(\ref{eq:wave_packet_transverse}), the momentum transfer is ${\bm Q} = {\bm p}_f - {\bm p}_i$ with ${\bm p_i} = \left< {\bm k}_i \right> = (0, 0, p_i)$ being an \textit{averaged} momentum of the incident electron, ${\bm p}_f$ the momentum of the scattered electron, and $f({\bm Q} - {\bm k}_\perp)$ is the \textit{plane--wave} scattering amplitude in the first--Born approximation:
\begin{equation}
   \label{eq:amplitude_plane_wave}
   f({\bm q}) = - \frac{m_e}{2 \pi} \, \int U(r) \, {\rm e}^{-i {\bm q} {\bm r}} \,
   {\rm d}^3{\bm r} \, .
\end{equation}
We have assumed here that the interaction of electrons with a target is described by a central potential $U(r)$ and that the outgoing (scattered) electrons are detected as plane--waves with the wave--vector ${\bm p}_f$. In Eq.~(\ref{eq:amplitude_monochromatic}), moreover, we have introduced the exponential factor ${\rm exp}(i {\bm k}_\perp {\bm b})$ in order to specify the lateral position of the scatterer with regard to the central ($z$--) axis of the incident wave--packet. Here, ${\bm b} = (b_x, b_y, 0)$ is the impact parameter which \textit{vanishes} when the potential is placed in the center of the beam.

Along with Eq.~(\ref{eq:amplitude_monochromatic}), one can derive another representation for the scattering amplitude $F({\bm Q}, {\bm b})$ that, in some cases,
can be more convenient for theoretical analysis. By inserting the standard plane--wave amplitude (\ref{eq:amplitude_plane_wave}) into Eq.~(\ref{eq:amplitude_monochromatic}) we find:
\begin{equation}
   \label{eq:amplitude_wave_packet}
   F({\bm Q}, {\bm b}) = - \frac{m_e}{2 \pi} \, \int U(r) \,
   \Psi_{\rm tr}^{(m)}({\bm r} + {\bm b}) \, {\rm e}^{-i {\bm Q} {\bm r}} \, {\rm d}{\bm r} \, ,
\end{equation}
where
\begin{equation}
   \label{eq:wave_packet_shifted}
   \Psi_{\rm tr}^{(m)}({\bm r} + {\bm b}) = \int {\rm e}^{i ({\bm r} + {\bm b}) {\bm k}_\perp} \,
   \Phi_{\rm tr}^{\rm (m)}({\bm k}_\perp) \, \frac{{\rm d}^2 {\bm k}_\perp}{2 \pi} \, .
\end{equation}
This expression can help to analyze (at least qualitatively) scattering at the small impact parameters. For $b = 0$, in particular, the function $U(r) \, \Psi_{\rm tr}^{(m)}({\bm r} + {\bm b}) = U(r) \, \Psi_{\rm tr}^{(m)}({\bm r})$ under the integral in Eq.~(\ref{eq:amplitude_wave_packet}) corresponds to a state with a definite projection of the orbital angular momentum $m$. The expansion of this function into spherical harmonics $Y_{lm}(\theta_r, \varphi_r)$ may contain, therefore, \textit{only} multipoles with the OAM $l \ge |m|$. It results in a specific behaviour of the angular distributions of scattered electrons that become most pronounced for the forward direction. Indeed, by substituting the transverse component (\ref{eq:transverse_component_convolution}) of the wave--packet into the scattering amplitude (\ref{eq:amplitude_wave_packet}), we obtain:
\begin{eqnarray}
   \label{eq:amplitude_wave_packet_b_0}
   F({\bm Q}, {\bm b} = 0) && \nonumber \\
   && \hspace*{-2.3cm} = - \frac{m_e}{2 \pi} \int r_\perp \, {\rm d}r_\perp {\rm d}z\,
   U\left(\sqrt{r_\perp^2 + z^2}\right) \, R^{(m)}\left(r_\perp \right) \,
   {\rm e}^{-iQ_z z} \nonumber \\
   &\times& \int_0^{2\pi} {\rm e}^{i m \varphi_r - iQ_\perp r_\perp
   \cos{(\varphi_r - \varphi)}} \, \frac{{\rm d}\varphi_r}{2\pi} \, ,
\end{eqnarray}
where $Q_\perp = p_{f, \perp} = p_f (\sin\theta \cos\varphi, \sin\theta \sin\varphi, 0)$, $Q_z = p_f \cos\theta - p_i$,  and the integral over $\varphi_r$ is reduced to:
\begin{eqnarray}
   \int_0^{2\pi} \, {\rm e}^{i m \varphi_r - i Q_\perp r_\perp \cos(\varphi_r - \varphi)}\,
   \frac{{\rm d}\varphi_r}{2 \pi} && \nonumber \\
   && \hspace*{-3cm} = {\rm e}^{i m \varphi} \, (-i)^m \, J_m(Q_\perp r_\perp) \,.
\end{eqnarray}
From these expressions and from the asymptotic behaviour of the Bessel function, $|J_m(x)| = (x/2)^{|m|}/|m|!$ for $x \to 0$, one finds that the amplitude
\begin{eqnarray}
   \label{eq:amplitude_wave_packet_b_0_forward}
   F({\bm Q}, {\bm b} = 0) \propto Q_{\perp}^{|m|} \propto (\sin\theta)^{|m|}
\end{eqnarray}
\textit{vanishes} for $\theta \to 0$ when the OAM projection $m \ne 0$. Since the angle--differential cross section is proportional to the square of $F({\bm Q}, {\bm b})$, the emission pattern of the scattered electrons develops a dip in the forward direction and for the central collision, $b = 0$. However behaviour of the function $U(r) \Psi_{\rm tr}^{(m)}(\bm r+\bm b)$ changes with the growth of $\bm b$, and the dip at $\theta=0$ disappears. This angular behaviour, peculiar to twisted electron beams, will be discussed in detail later.

A further analysis of $F({\bm Q}, {\bm b})$ requires knowledge of the interaction potential $U(r)$. In Section \ref{sec:Yukawa_and_hydrogenic_formulas}, for example, we will show how this scattering amplitude is calculated for the Yukawa- and hydrogenic potentials.

%
%
\subsection{Number of events and cross sections}
\label{subsec:cross_sections_number_of_events}

With the help of the amplitude $F({\bm Q}, {\bm b})$ one can calculate the number of events and the cross section for potential scattering of a twisted electron's wave--packet. The explicit form of these observables depends on a set--up of the scattering ``experiment'' and, in particular, on the composition of the target. In this section, we study scattering off (i) a single potential,
(ii) the infinitely wide (macroscopic) targets, and (iii) off the localized (mesoscopic) targets.

%
%
\subsubsection{Scattering by a single potential}
\label{subsubsect:single_potential}

We start the discussion with a \textit{single} potential $U(r)$ located at the impact parameter ${\bm b}$ from the center of the electron wave--packet. As discussed already in Ref.~\cite{KaK16}, a usual definition of the cross section is not applicable in this case. Instead, the scattering process can be described by a number of events:
\begin{equation}
	\label{eq:numebr_events_single_potential_general}
	\frac{{\rm d}\nu}{{\rm d}\Omega}=\frac{N_e}{\cos{\theta_k}} \,
    \left| F({\bm Q}, {\bm b}) \right|^2 \, ,
\end{equation}
where $N_e$ is the number of electrons in the incident beam.

Eq.~(\ref{eq:numebr_events_single_potential_general}) can be further simplified for the so--called \textit{wide} wave--packet. Here, the distribution function $g_{\varkappa_0 \sigma_\varkappa}(k_\perp)$ from Eq.~(\ref{eq:transverse_component_convolution}) is sharply peaked at $\left< k_\perp \right> = \varkappa_0 \ne 0$, so that $1/\sigma_\varkappa \gg a$. In this case, the transverse momentum ${\bm k}_\perp = k_\perp \left(\cos{\varphi_k}, \sin{\varphi_k}, 0 \right)$ of the incident electron wave can be approximated as:
\begin{equation}
   \label{eq:k_perp_0}
   {\bm k}^{(0)}_\perp= \varkappa_0 \,
   \left(\cos{\varphi_k}, \sin{\varphi_k}, 0 \right) \, .
\end{equation}
By substituting ${\bm k}_\perp \to {\bm k}^{(0)}_\perp$ in the plane--wave amplitude $f({\bm Q} - {\bm k}_\perp)$ and making use of Eqs.~(\ref{eq:amplitude_monochromatic}) and (\ref{eq:numebr_events_single_potential_general}), we find:
\begin{equation}
	\label{eq:eq:numebr_events_single_potential_general_wide}
	\frac{{\rm d}\nu}{{\rm d}\Omega} = L^{({\rm tw})} \,
	\left| \int_0^{2\pi} \, f\left({\bm Q} - {\bm k}^{(0)}_\perp \right) \,
	{\rm e}^{i m\varphi_k + i{\bm k}^{(0)}_\perp {\bm b}} \,
	\frac{{\rm d}\varphi_k}{2\pi} \right|^2 \, ,
\end{equation}
where the quantity
\begin{equation}
	\label{eq:Ltw_general}
	L^{({\rm tw})} = \frac{N_e}{\cos{\theta_k}} \,
	\left|\int_0^\infty g_{\varkappa_0 \sigma_\varkappa}(k_\perp) \sqrt{\frac{k_\perp}{2\pi}} \, {\rm d}k_\perp \right|^2
	\,
\end{equation}
can be viewed as luminosity of the collision.

When comparing this with Eq.~(\ref{eq:transverse_component_convolution}), one sees that $L^{({\rm tw})}$ can be expressed in terms of the transverse density of the incident Bessel packet with $m = 0$ at the coordinate origin:
\begin{equation}
	\label{eq:Ltw_general_PSI}
	L^{({\rm tw})} = \frac{N_e}{\cos{\theta_k}} \,
	\left|\Psi^{(0)}_{\rm tr}({\bm r}_\perp={\bf 0}) \right|^2\,.
\end{equation}
Together with Eq.~(\ref{eq:eq:numebr_events_single_potential_general_wide}), this expression indicates that for $m = 0$ and $\theta_k \to 0$ we obtain the number of events for the ``standard'' plane--wave case, see Eq.~(4) from Ref.~\cite{KaK16}. Therefore, the quantity
 \begin{equation}
 \frac{1}{L^{\rm (tw)}}\, \frac{{\rm d}\nu}{{\rm d}\Omega}\equiv \frac{d\sigma^{\rm (tw)}(\bm b)}{d\Omega}
 \label{cross-single}
 \end{equation}
can be considered as a cross section for a single potential, localized at the impact parame\-ter~$\bm b$. In particular, at $\varkappa_0 \ll Q$ we obtain the following simple formula:
 \begin{equation}
 \frac{{\rm d}\sigma^{\rm (tw)}(\bm b)}{{\rm d}\Omega}=
 \frac{{\rm d}\sigma^{\rm (PW)}}{{\rm d}\Omega}\,
 J_m^2(\varkappa_0 b).
 \label{cross-single-up}
 \end{equation}
%

%
%
\subsubsection{Infinitely wide (macroscopic) target}
\label{subsubsec_wide_target}

After discussion of the scattering by a single potential, let us now turn to the \textit{macroscopic} target, which consists of randomly distributed force centers and has a transverse extension ${\mathcal R} \gg 1/\sigma_{\varkappa}$. For such a target one can introduce the \textit{averaged} cross section:
\begin{eqnarray}
   \label{eq:cross_section_macroscopic_target}
   \frac{{\rm d}{\bar \sigma}}{\rm{d}\Omega} &=& \frac{1}{\cos\theta_k} \,
   \int \left| F({\bm Q}, {\bm b}) \right|^2 \, {\rm d}^2{\bm b}  \\
   && \hspace*{-1cm} = \frac{1}{\cos\theta_k} \, \int_0^{2\pi} \frac{{\rm d}\varphi_k}{2\pi} \,
   \int_0^\infty {\rm d}k_\perp \left| g_{\varkappa_0 \sigma_\perp}(k_\perp)\right|^2 \,
   \left|f({\bm Q} - {\bm k}_\perp)\right|^2 \,  \nonumber
\end{eqnarray}
which is obtained (i) by integrating the number of events over all the impact parameters $b$, and (ii) by normalizing the result by a number of incident electrons, see Eqs. (33) and (38) in Ref.~\cite{KaK16}.
As can be seen from Eq.~({\ref{eq:cross_section_macroscopic_target}), ${\rm d}{\bar \sigma}/{\rm d}\Omega$ depends neither on the projection $m$ of the orbital angular momentum nor on the spatial structure of an incident phase front.

One can further simplify Eq.~(\ref{eq:cross_section_macroscopic_target}) in the approximation of a wide wave--packet. As mentioned already, the momentum distribution function $g_{\varkappa_0 \sigma_\varkappa}(k_\perp)$ is sharply peaked in this case at $\left< k_\perp \right> = \varkappa_0$ and the wave--packet (\ref{eq:wavepacket_Bessel_coordinate}) approaches the monochromatic limit. Similar to before, one can approximate the transverse momentum of the incident electron ${\bm k}_\perp$ by ${\bm k}_{\perp}^{(0)}$, given by Eq.~(\ref{eq:k_perp_0}). By substituting ${\bm k}_\perp = {\bm k}_{\perp}^{(0)}$ in Eq.~(\ref{eq:cross_section_macroscopic_target}) and taking the square of the (plane--wave) scattering amplitude $f({\bm Q} - {\bm k}^{(0)}_\perp)$ out of the integral over $k_\perp$, we obtain:
\begin{equation}
   \label{eq:cross_section_macroscopic_target_wide_packet}
   \frac{{\rm d}{\bar \sigma}}{\rm{d}\Omega} =
   \frac{1}{\cos\theta_k} \, \int_0^{2\pi} \frac{{\rm d}\varphi_k}{2\pi} \,
   \left|f({\bm Q} - {\bm k}^{(0)}_\perp)\right|^2 \, ,
\end{equation}
where, as usual, the momentum transfer is ${\bm Q} = {\bm p}_f - {\bm p}_i$, and $|{\bm p}_f|=\left|{\bm p}_i + {\bm k}^{(0)}_\perp\right|$.

As can be seen from Eq.~(\ref{eq:cross_section_macroscopic_target_wide_packet}), the calculation of the scattering cross section for a wide wave--packet and a macroscopic target can be traced back to the plane--wave amplitude $f({\bm Q} - {\bm k}^{(0)}_\perp)$. Therefore, based on the  properties of this amplitude we can predict the main features of ${\rm d}{\bar \sigma}/\rm{d}\Omega$. It is well--known, for example, that for Yukawa and hydrogenic potentials  $f({\bm q})$ depends solely on the ${\bm q}^2$, see Ref.~\cite{KaK16}. Since in Eq.~(\ref{eq:cross_section_macroscopic_target_wide_packet}) the argument of the plane--wave scattering amplitude is ${\bm Q} - {\bm k}^{(0)}_\perp$ and the square of this momentum transfer is given by:
\begin{eqnarray}
   \label{eq:Q2_formula}
   \left({\bm Q} - {\bm k}^{(0)}_\perp \right)^2 &&\\
   && \hspace*{-2cm} = 2p^2_f\,\left[1-\cos\theta \cos\theta_k-\sin\theta\sin\theta_k \cos(\varphi_k-\varphi)\right] \, , \nonumber
\end{eqnarray}
we can re--write Eq.~(\ref{eq:cross_section_macroscopic_target_wide_packet}) as follows:
\begin{equation}
\label{eq:cross_section_macroscopic_target_wide_packet_2}
\frac{{\rm d}{\bar \sigma}(\theta; \theta_k)}{\rm{d}\Omega} =
\frac{1}{\cos\theta_k} \, \int_0^{2\pi} \frac{{\rm d}(\varphi_k - \varphi)}{2\pi} \,
\left|f({\bm Q} - {\bm k}^{(0)}_\perp)\right|^2 \, ,
\end{equation}
where $\varphi$ is the azimuthal angle of outgoing electrons. Eqs.~(\ref{eq:Q2_formula})--(\ref{eq:cross_section_macroscopic_target_wide_packet_2}) reveal that the scattering pattern is azimuthally symmetric with respect to the beam ($z$--) axis and is peaked at $\theta = \theta_k$. The latter follows also from the fact that $f({\bm Q} - {\bm k}^{(0)}_\perp)$ is maximal at $\left({\bm Q} - {\bm k}^{(0)}_\perp\right)^2 = 0$ for the Yukawa and hydrogenic potentials, see Ref.~\cite{KaK16}.

Eq.~(\ref{eq:cross_section_macroscopic_target_wide_packet_2}) can now be used to calculate the total cross section. Namely, by integrating over the directions of outgoing electrons, we find:
\begin{eqnarray}
   \label{eq:total_cross_section_wide_packet}
   \bar \sigma &=& \frac{1}{\cos{\theta_k}}
   \int_0^{2\pi} \frac{{\rm d}\varphi_k}{2\pi}
   \int {\rm d}\Omega \left|f\left({\bm p}_f- {\bm p}_i - {\bm k}^{(0)}_\perp\right)\right|^2
   \nonumber \\
   &=& \frac{\sigma_{\rm pl}}{\cos{\theta_k}} \, ,
\end{eqnarray}
where $\sigma_{\rm pl}$ is the plane--wave cross section. Clearly, the cross section for the scattering of twisted electrons by a macroscopic target is generally \textit{larger} than $\sigma_{\rm pl}$.

\subsubsection{Superposition of two twisted beams}
\label{subsubsec_two_twisted_beams}

As we have seen from Eqs.~(\ref{eq:cross_section_macroscopic_target}) and (\ref{eq:cross_section_macroscopic_target_wide_packet_2}), the angle--differential cross section for the scattering by a macroscopic (infinite) target is independent of the projection of the orbital angular momentum $m$ and of the phase structure of the incident twisted beam. Later, in Section \ref{subsubsec:scattering_localized}, we will show that this OAM's-- and phase--sensitivity is restored for spatially localized scatterers and well--focused electron beams. However, even in the ``large--target---wide--packet'' regime the averaged cross section can be sensitive to the OAM if electrons are initially prepared in a coherent superposition of two states with the same kinematic parameters but different $m$:
\begin{equation}
   \label{eq:superposition_two_states}
   \Psi_{\varkappa_0 p_i}({\bm r}) = c_1 \Psi_{\varkappa_0 m_1 p_i}({\bm r}) + c_2 \Psi_{\varkappa_0 m_2 p_i}({\bm r})
\end{equation}
where $\Psi_{\varkappa_0 m p_i}({\bm r})$ is given by Eq.~(\ref{eq:wavepacket_Bessel_coordinate}) and the expansion coefficients are:
\begin{eqnarray}
   \label{eq:c_coefficients}
   c_n = \left| c_n \right| \, {\rm e}^{i \alpha_n}, \; \; \; \left|c_1\right|^2 + \left|c_2\right|^2 = 1 \, .
\end{eqnarray}
By inserting the superposition (\ref{eq:superposition_two_states}) into Eqs.~(\ref{eq:amplitude_monochromatic}) and (\ref{eq:cross_section_macroscopic_target}), and passing to the limit $\sigma_\varkappa \to 0$ we obtain:
\begin{eqnarray}
   \label{eq:cross_section_macroscopic_target_wide_packet_superposition_1}
   \frac{{\rm d}{\bar \sigma}^{\rm (2)}(\theta, \varphi ; \, \theta_k)}{\rm{d}\Omega} &=& \frac{1}{\cos\theta_k}
   \nonumber \\
   && \hspace*{-3.0cm} \times
   \int_0^{2\pi} \frac{{\rm d}\varphi_k}{2\pi} \,
   \left|f({\bm Q} - {\bm k}^{(0)}_\perp)\right|^2 \, G(\varphi_k, \Delta m, \Delta \alpha) \, ,
\end{eqnarray}
with $\Delta m = m_2 - m_1$, $\Delta \alpha = \alpha_2 - \alpha_1$ and the factor:
\begin{eqnarray}
   \label{eq:G_factor}
   G(\varphi_k, \Delta m, \Delta \alpha) &=& 1 + 2 \left|c_1 c_2 \right| \nonumber \\
   && \hspace*{-2cm} \times \cos\Big[(m_2-m_1)(\varphi_k-\pi/2) + \alpha_2 - \alpha_1 \Big].
\end{eqnarray}
In order to further evaluate this expression, we note that the plane--wave scattering amplitude $f({\bm Q} - {\bm k}^{(0)}_\perp)$ depends on the difference between the azimuthal angles $\varphi_k - \varphi$, see Eq.~(\ref{eq:Q2_formula}). This angular dependence allows us to perform an integration over $\varphi_k$ and to find:
\begin{eqnarray}
   \label{eq:cross_section_macroscopic_target_wide_packet_superposition_2}
   \frac{{\rm d}{\bar \sigma}^{\rm (2)}(\theta, \varphi ; \, \theta_k)}{\rm{d}\Omega} &=&
   \frac{{\rm d}{\bar \sigma}(\theta ; \, \theta_k)}{\rm{d}\Omega}
   \nonumber \\[0.2cm]
   && \hspace*{-2.0cm} \times \Big[1 + A(\theta; \theta_k) \,
   \cos{[\Delta m(\varphi-\pi/2)+\Delta\alpha]} \Big] \, ,
\end{eqnarray}
where ${\rm d}{\bar \sigma}/\rm{d}\Omega$ is the cross section (\ref{eq:cross_section_macroscopic_target_wide_packet_2}) for a single Bessel beam, and the azimuthal asymmetry parameter reads as follows:
\begin{eqnarray}
   \label{eq:cross_section_macroscopic_target_wide_packet_superposition_interference}
   A(\theta; \theta_k) &=& \left( \frac{{\rm d}{\bar \sigma}(\theta ; \, \theta_k)}{\rm{d}\Omega} \right)^{-1} \,
   \frac{2 \left|c_1 c_2\right|}{\cos\theta_k}  \\
   && \hspace*{-2cm} \times \int_0^{2\pi} \,
   \left| f({\bm Q} - {\bm k}^{(0)}_\perp)\right |^2 \,
   \cos{[\Delta m(\varphi_k-\varphi)]}\frac{{\rm d}(\varphi_k-\varphi)}{2\pi} \, . \nonumber
\end{eqnarray}
These expressions indicate that the angle--differential cross section ${\rm d}{\bar \sigma}^{\rm (2)}/\rm{d}\Omega$ exhibits an azimuthal asymmetry which depends on the difference of the OAM's projections $\Delta m = m_2 - m_1$ and of the phases $\Delta \alpha = \alpha_2 - \alpha_1$. This dependence, however, does not appear in the total cross section:
\begin{eqnarray}
   \label{eq:total_cross_section_wide_packet_two beams}
   {\bar \sigma}^{\rm (2)} &=& \int \frac{{\rm d}{\bar \sigma}^{\rm (2)}(\theta, \varphi ; \, \theta_k)}{\rm{d}\Omega} \, {\rm d}\Omega = \frac{\sigma^{\rm (pl)}}{\cos\theta_k} \, ,
\end{eqnarray}
since the second term in Eq.~(\ref{eq:cross_section_macroscopic_target_wide_packet_superposition_2}) vanishes identically after the integration over the scattering angles.

%
%
\subsubsection{Scattering by a localized (mesoscopic) target}
\label{subsubsec:scattering_localized}

Until now we have discussed the potential scattering for two extreme cases of either a single--potential or a macroscopic (infinitely wide) target. In a more realistic experimental scenario, a focused electron beam collides with a well--localized mesoscopic atomic target. In order to account for geometrical effects in such a scenario, we describe a target as an incoherent ensemble of potential centers. The density of the scatterers in the transverse ($xy$--) plane is characterized by a distribution function $n({\bm b})$, which is normalized as follows:
\begin{equation}
\label{eq:electron_distribution}
\int n({\bm b}) \, {\rm d}^2{\bm b} = 1 \, .
\end{equation}
With the help of this distribution, we can also obtain a form factor of the mesoscopic target:
\begin{equation}
   \label{eq:form_factor_target}
   {\mathcal F}({\bm k}_\perp) = \int n({\bm b}) \, {\rm e}^{-i {\bm k}_\perp {\bm b}} \,
   {\rm d}^2{\bm b} \, ,
\end{equation}
which later will be employed to simplify expression for the number of scattering events.

By making use of Eqs.~(29)--(30) from Ref.~\cite{KaK16} we find the number of scattering events:
\begin{equation}
   \label{eq:number_of_events_localized_target}
   \frac{{\rm d} \nu}{{\rm d} \Omega}= \frac{N_e}{\cos{\theta_k}} \,
   \int \left| F({\bm Q}, {\bm b}) \right|^2 \,
   n({\bm b}) \, {\rm d}^2{\bm b} \, ,
\end{equation}
where $N_e$ is the total number of electrons in the incident twisted bunch.
To perform an integration over the impact parameter ${\bm b}$ we insert into this expression the amplitude $F({\bm Q}, {\bm b})$ from (\ref{eq:amplitude_monochromatic}):
\begin{eqnarray}
   \label{eq:number_of_events_localized_target_2}
   \frac{{\rm d} \nu}{{\rm d} \Omega} &=& \frac{N_e}{\cos{\theta_k}} \, \int {\rm d}^2{\bm b}
   \, n({\bm b}) \, \Bigg[ \int \frac{{\rm d}^2{\bm k}_\perp}{2 \pi} \, f({\bm Q} - {\bm k}_\perp) \,
   \Phi^{(m)}_{\rm tr}({\bm k}_\perp) \nonumber \\[0.2cm]
   &\times& \int \frac{{\rm d}^2{\bm k}'_\perp}{2 \pi} \, f^*({\bm Q} - {\bm k}'_\perp) \,
   \Phi^{(m) *}_{\rm tr}({\bm k}'_\perp) \, {\rm e}^{i ({\bm k}_\perp-{\bm k}'_\perp) {\bm b}}\Bigg].
\end{eqnarray}
Using the explicit form of the transverse component $\Phi^{(m)}_{\rm tr}({\bm k}_\perp)$, as given by Eq.~(\ref{eq:wave_packet_transverse}), and noting that

\begin{equation}
   \label{eq:b_integral_target_density}
   \int {\rm e}^{i({\bm k}_\perp - {\bm k}'_\perp) {\bm b}} \, n({\bm b}) \,{\rm d}^2{\bm b} =
   {\mathcal F}({\bm k}'_\perp - {\bm k}_\perp) \, ,
\end{equation}
we arrive at the following formula:
\begin{eqnarray}
   \label{eq:number_of_events_localized_target_3}
   \frac{{\rm d} \nu}{{\rm d} \Omega} &=& \frac{N_e}{\cos{\theta_k}} \int
   \frac{{\rm d}^2{\bm k}_\perp}{2 \pi} \, \frac{{\rm d}^2{\bm k}'_\perp}{2 \pi} \,
   f({\bm Q} - {\bm k}_\perp) \, f^*({\bm Q} - {\bm k}'_\perp) \nonumber \\
   && \hspace*{-1cm} \times{\mathcal F}({\bm k}'_\perp - {\bm k}_\perp) \,
   \frac{{\rm e}^{i m (\varphi_k - \varphi_{k'})}}{2\pi \sqrt{k_\perp k'_\perp}}
   g_{\varkappa_0 \sigma_\varkappa}(k_\perp) g_{\varkappa_0 \sigma_\varkappa}(k'_\perp) \, .
\end{eqnarray}
Further evaluation of this number of events requires the knowledge of (i) the target density $n({\bm b})$ as well as of (ii) the plane--wave amplitude $f({\bm Q} - {\bm k}_\perp)$.
For the numerical analysis below we take $n({\bm b})$ to be a Gaussian function:
\begin{equation}
   \label{eq:Gaussian_distribution_2}
   n({\bm b}) = \frac{1}{2 \pi \sigma_b^2} \, {\rm e}^{-\frac{({\bm b} - {\bm b}_0)^2}{2 \sigma_b^2}}
   \, .
\end{equation}
This distribution is sharply peaked at the impact parameter ${\bm b}_0$ and its form factor reads as:
\begin{equation}
	\label{eq:Gaussian_form_factor_2}
	{\mathcal F}({\bm k}_\perp) = {\rm e}^{-i {\bm k}_\perp {\bm b}_0 -
    {\bm k}^2_\perp \sigma_b^2/2}\, .
\end{equation}

Eq.~(\ref{eq:number_of_events_localized_target_3}) can be further simplified for two limiting cases. When the target is \textit{considerably smaller} than the incident wave-packet,
 \be
 \sigma_b \ll 1/\sigma_\varkappa,
 \ee
we get
 \be
 \frac{{\rm d} \nu}{{\rm d} \Omega} =
 L^{\rm (tw)}\, \fr{{\rm d}\sigma^{\rm (mesos)}}{{\rm d}\Omega},
 \ee
where 
 \bea
 \fr{{\rm d}\sigma^{\rm (mesos)}}{{\rm d}\Omega}
 &=& \int_0^{2\pi} \fr{d\varphi_k} {2\pi}\fr{d\varphi_{k'}}{2\pi}
 f\left(\bm Q-\bm k_\perp^{(0)}\right)\,f^*\left(\bm Q-\bm k_\perp^{'(0)}\right)
  \nn
  \\
  &&\times {\rm e}^{im(\varphi_k-\varphi_{k'})}\,
 {\cal F}\left(\bm k_\perp^{'(0)}-\bm k_\perp^{(0)}\right).
 \nn
 \eea
If, additionally, the inequality $\varkappa_0 \ll Q$ is fulfilled, then the cross section becomes
 \be
 \fr{d\sigma^{\rm (mesos)}}{d\Omega}=\fr{d\sigma^{\rm (PW)}}{d\Omega}\,
 \int J_m^2(\varkappa_0 b)\, n(\bm b) \,d^2 \bm b,
 \ee
which results in a one-dimension integral for the Gaussian distribution~(\ref{eq:Gaussian_distribution_2}) of atoms in the target:
  \bea
  \label{ratioRb0}
 &&\fr{d\sigma^{\rm (mesos)}(b_0)}{d\Omega}= R(b_0)\, \fr{d\sigma^{\rm (PW)}}{d\Omega}\,,
  \\
 && R(b_0)=\int_0^\infty J^2_m(\varkappa_0 b) I_0(bb_0/\sigma^2_b)\,{\rm e}^{-(b^2+b_0^2)/(2\sigma^2_b)}\, \fr{b\,db}{\sigma^2_b}.
   \nn
  \eea

When the target is \textit{considerably larger} than the incident wave-packet,
 \be
 \sigma_b \gg 1/\sigma_\varkappa,
 \ee
and the amplitude $f$ is a real function ($\text{Arg}\,f = 0$), we obtain
 \bea
 \fr{d\nu}{d\Omega}&=&\fr{N_e}{\cos\theta_k}
 \int\fr{d^2 \bm k_\perp}{2\pi k_\perp}\, g^2(k_\perp)\,
   \left|f\left(\bm Q-\bm k_\perp\right) \right|^2 \,
   n\left(\bm b_k \right),
   \nn
   \\
 \bm b_k&=&\fr{m}{k_\perp}\,(\sin\varphi_{k}\,, -\cos\varphi_{k}\,, 0).
 \label{eq:large-beam}
  \eea

The scattering amplitude $f({\bm Q} - {\bm k}_\perp)$ depends on the potential $U(r)$ and it is real in the examples of Yukawa and hydrogenic potentials given in the next Section.

%
%
\section{Scattering by Yukawa and hydrogenic potentials}
\label{sec:Yukawa_and_hydrogenic_formulas}

In Section \ref{sec:theory} we have derived the number of events and the cross sections for scattering of the Bessel electron's wave--packets by a single--potential and by the mesoscopic-- and macroscopic targets. In that analysis, however, a particular form of the central potential $U(r)$ was not specified. Below we consider two specific cases, when $U(r)$ is the Yukawa-- and hydrogenic potential,
and discuss properties of the scattered electrons.

%
%
\subsection{Yukawa potential}
\label{susec:Yukawa_potential}

The Yukawa potential
\begin{equation}
   \label{eq:Yukawa_potential}
   U(r) = \frac{V_0}{r} \, {\rm e}^{-\mu r}
\end{equation}
describes the field with a typical radius of action $a\sim 1/\mu$. In a theory of atomic collisions, it is used very often as an approximation to the Coulomb field of the nucleus screened by atomic electrons, see for example Ref.~\cite{Sal91,SaM87}. This potential allows one to evaluate analytically the standard plane--wave scattering amplitude within the first--Born approximation:
\begin{eqnarray}
   \label{eq:amplitude_plane_wave_Yukawa}
   f({\bm q}) = \frac{-2 m_e V_0}{q^2 + \mu^2} \, .
\end{eqnarray}
By inserting this expression into Eq.~(\ref{eq:amplitude_monochromatic}) and making simple algebra we derive the scattering amplitude for the twisted electron:
\begin{eqnarray}
   \label{eq:amplitude_wave_packet_twisted_Yukawa}
   F({\bm Q}, {\bm b}) &=& - \frac{(-i)^m \, 2 m_e \, V_0 \, {\rm e}^{i m \varphi}}{\sqrt{2\pi}} \nonumber \\[0.1cm]
   && \hspace*{-1.5cm} \times \int_0^\infty {\rm d}k_\perp \, g_{\varkappa_0 \sigma_\varkappa}(k_\perp) \, \sqrt{k_\perp} \, I_{m}(\alpha, \beta, {\bm b}) \, ,
\end{eqnarray}
where the function
\begin{equation}
	\label{eq:I_function}
	I_{m}(\alpha, \beta, {\bm b}) = \int_0^{2\pi}
	\frac{{\rm d}\psi}{2\pi} \;
	\frac{{\rm e}^{im\psi+ik_\perp b\cos(\psi + \varphi-\varphi_b)}}{\alpha-\beta\,\cos{\psi}}
\end{equation}
was introduced and studied in Ref.~\cite{SeI15}. Here both $\alpha$ and $\beta$ are positive:
\begin{eqnarray}
	\label{eq:alpha_beta}
	\alpha &=& Q^2 + k_\perp^2 + \mu^2 \nonumber \\
	&=& p^2_f(1+\cos^2{\theta_k} - 2\cos{\theta}\cos{\theta_k})+k^2_\perp+\mu^2 \, ,
	\nonumber \\[0.2cm]
	\beta &=& 2 k_\perp Q_\perp =2k_\perp p_f \sin{\theta} \, ,
\end{eqnarray}
and $\alpha > \beta$. By making use of Eqs.~(\ref{eq:amplitude_wave_packet_twisted_Yukawa})--(\ref{eq:alpha_beta}) one can study the potential scattering of a twisted electron for any experimental ``scenario''. Below we treat a few such scenarios.

First, let us study collision with a single Yukawa scatterer, located at the central axis of the incident packet.
For such a \textit{central collision} the impact parameter ${\bm b} = 0$ and the function (\ref{eq:I_function}) simplifies to:
\begin{equation}
   \label{eq:I_function_central_collision}
   I_m(\alpha, \beta, 0) =
   \left( \frac{\beta}{\alpha + \sqrt{\alpha^2-\beta^2}} \right)^{|m|}
   \frac{1}{\sqrt{\alpha^2-\beta^2}} \, ,
\end{equation}
see Ref.~\cite{SeI15} for further details. With the help of this expression we can study angular distribution of the scattered electrons.
The small scattering angles are of specific interest here, for which $I_m(\alpha, \beta, 0) \propto \beta^{|m|} \propto (\sin \theta)^{|m|}$. Such a $\theta$--behaviour implies that for $m \ne 0$ the angular distribution
\begin{equation}
   \label{eq:angular_distribution_single_Yukawa}
   \frac{{\rm d}\nu}{{\rm d}\Omega} \propto (\sin{\theta})^{2|m|}
   \; \; {\rm for} \; \; \theta \to 0 \, ,
\end{equation}
vanishes for the forward emission, $\theta \to 0$. We remind the reader that this result was predicted above on a basis of the analysis of the scattering amplitude, see Eq.~(\ref{eq:amplitude_wave_packet_b_0_forward}).

The dip in the scattering pattern at $\theta \to 0$ disappears, however, with the increase of the impact parameter $b$. Indeed, since for the forward scattering the function (\ref{eq:I_function}) reads as
\begin{equation}
	\label{eq:I_function_forward_emission}
	 I_m(\alpha, \beta, \bm b){\Large |}_{\theta=0}= \frac{e^{-im(\varphi-
    \varphi_b+\pi/2)}}{\alpha}
	 \;J_m(k_\perp b) \, ,	
\end{equation}
and $J_m(k_\perp b) \propto b^{|m|}$ for small impact parameters, we find:
\begin{equation}
\label{eq:angular_distribution_single_Yukawa_2}
\frac{{\rm d}\nu}{{\rm d}\Omega} \propto b^{2|m|} \; \; {\rm for} \; \; b \to 0 \, .
\end{equation}
This expression predicts that the (forward) electron emission quickly increases if the scattering center is shifted from the central beam axis.

In the second scenario, we still focus on the scattering off a single potential but for a particular case of a \textit{wide} wave--packet. The number of events for this scenario can be obtained from Eq.~(\ref{eq:eq:numebr_events_single_potential_general_wide}) as follows:
\begin{equation}
   \label{eq:number_of_events_Yukawa_wide_packet}
    \frac{{\rm d}\nu}{{\rm d}\Omega} =
    L^{({\rm tw})} \, \left| 2 m_e V_0 \, I_m(\alpha_0, \beta_0, b) \right|^2 \, ,
\end{equation}
where $\alpha_0$ and $\beta_0$ are given by Eq.~(\ref{eq:alpha_beta}) in which $k_\perp = \varkappa_0$ and $L^{({\rm tw})}$ is from Eq.~(\ref{eq:Ltw_general}).
This expression can be further simplified for the central collision, that is, for $b = 0$:
\begin{equation}
	\label{eq:number_of_events_Yukawa_wide_packet_central}
	\frac{{\rm d}\nu}{{\rm d} \Omega}= L^{({\rm tw})} \, \left|f_V^B \right|^2 \,,
\end{equation}
where we made use of Eq.~(\ref{eq:I_function_central_collision}) and introduced the ``amplitude'':
\begin{eqnarray}
	\label{eq:fBV_amplitude}
	f_V^B &=& -(-i)^m \, 2 m_e V_0 \, {\rm e}^{i m\varphi} \,
	\left(\frac{v}{u+\sqrt{u^2-v^2}}\right)^{|m|} \nonumber \\
	&\times& \frac{1}{\sqrt{u^2-v^2}} \, ,
\end{eqnarray}
with $u= Q^2 + \varkappa_0^2 + \mu^2$ and $v = 2 \varkappa_0 Q_\perp$. Up to unessential pre--factor this amplitude coincides with $f_V^B$ reported in Eq.~(24) of Ref.~\cite{BoP14}. In that work, however, no indication was given of how $f_V^B$ is related to the number of events or to the cross section.

Up to now we have discussed the potential scattering off a single Yukawa potential. If, in contrast, a wide wave--packet collides with an infinitely extended target consisting of randomly distributed Yukawa scatterers, one can use Eq.~(\ref{eq:cross_section_macroscopic_target_wide_packet}) to derive the cross section:
\begin{eqnarray}
	\label{eq:cross_section_macroscopic_target_wide_packet_Yukawa}
	\frac{{\rm d}\bar\sigma(\theta; \theta_k)}{{\rm d}\Omega}
	&& \nonumber \\
	&& \hspace*{-1.5cm} = \frac{(2 m_e V_0)^2}{\cos{\theta_k}}
	\,\int_0^{2\pi}\frac{{\rm d}\varphi_k}{2\pi}
	\frac{1}{\left[u-v\cos{(\varphi_k-\varphi)}\right]^2}
	\nonumber \\[0.2cm]
	&& \hspace*{-1.5cm} =
    \frac{(2 m_e V_0)^2}{\cos{\theta_k}} \frac{u}{\sqrt{(u^2-v^2)^3}} \, .
\end{eqnarray}
Here, we used the same short--hand notations $u$ and $v$ as in Eq.~(\ref{eq:fBV_amplitude}). One can further modify this expression to account for the incident electron beam, prepared as a coherent superposition of two Bessel states, see Eq.~(\ref{eq:superposition_two_states}). In this case, the differential cross section ${\rm d}\bar\sigma^{(2)}(\theta, \varphi; \theta_k)/{\rm d}\Omega$ is given by the second line of Eq.~(\ref{eq:cross_section_macroscopic_target_wide_packet_Yukawa}) in which the additional factor (\ref{eq:G_factor}) should be inserted under the integral over $\varphi_k$.

%
%
\subsection{Hydrogen atom in its ground state}
\label{susec:hydrogenic_potential}

We have discussed above the scattering of twisted electrons by a Yukawa potential. A superposition of few such potentials can be employed to reproduce accurately the (static) potential of \textit{any} neutral atom \cite{Sal91,SaM87}. In this section, however, we study a particular case of atomic hydrogen in the ground $1s$ state. Its potential is represented as follows:
\begin{equation}
   \label{eq:hydrogenic_potential}
   U_{H}(r) = -\frac{e^2}{r} \,\left( 1+ \frac{r}{a_0} \right) \, {\rm e}^{-2r/a_0} \, ,
\end{equation}
where $a_0$ is the Bohr radius. By inserting this expression into Eqs.~(\ref{eq:amplitude_plane_wave}) and (\ref{eq:amplitude_monochromatic}), we find the scattering amplitudes for the incident plane--wave:
\begin{equation}
   \label{eq:amplitude_plane_wave_hydrogenic}
   f({\bm q}) = \frac{a_0}{2} \, \left(\frac{1}{1 + (q a_0 /2)^2} +
  \frac{1}{\left(1 + (q a_0 /2)^2 \right)^2} \right) \, ,
\end{equation}
and for the twisted electron's wave--packet:
\begin{eqnarray}
	\label{eq:amplitude_wave_packet_twisted_hydrogenic}
    F({\bm Q}, {\bm b}) &=& \frac{(-i)^m a_0}{2\sqrt{2\pi}} \,
    \int_0^\infty {\rm d}k_\perp \, g_{\varkappa_0 \sigma_\perp}(k_\perp) \sqrt{k_\perp} \nonumber \\
    && \hspace*{-1.5cm} \times \int_0^{2\pi} \frac{{\rm d} \varphi_k}{2\pi} \;
    \left(\frac{1}{w}+\frac{1}{w^2} \right) \; {\rm e}^{i m\varphi_k + ik_\perp b \cos{(\varphi_k-\varphi_b)}} \, ,
\end{eqnarray}
where we introduced the following notations:
\begin{eqnarray}
	\label{eq:w_alpha_beta_hydrogenic}
	w &=& \alpha - \beta \, \cos{(\varphi_k-\varphi)} \, , \nonumber \\[0.1cm]
	\alpha &=& 1 + \mbox{$\frac 14$} a_0^2 p^2_f \left[1 + \cos^2{\theta_k} - 2\cos{\theta} \cos{\theta_k}\right] + \mbox{$\frac 14$} a_0^2 k^2_\perp \, , \nonumber \\[0.1cm]
	\beta&=&\mbox{$\frac 12$} a_0^2 k_\perp p_f \sin{\theta} \, .
\end{eqnarray}
In order to simplify $F({\bm Q}, {\bm b})$ further one can use the identity:
\begin{equation}
	\label{eq:w_identity}
	\frac{1}{w} + \frac{1}{w^2}=
	\left(1 -\frac{\partial}{\partial \alpha} \right)\frac{1}{w} \, ,
\end{equation}
and then we re--write Eq.~(\ref{eq:amplitude_wave_packet_twisted_hydrogenic}) as follows:
\begin{eqnarray}
	\label{eq:amplitude_wave_packet_twisted_hydrogenic_2}
	F({\bm Q}, {\bm b}) &=& 	\frac{(-i)^m \,a_0\, {\rm e}^{im\varphi}}{2 \sqrt{2\pi}}
	\nonumber \\
	&& \hspace*{-2cm} \times
	\int_0^\infty {\rm d}k_\perp \, g_{\varkappa_0 \sigma_\perp}(k_\perp)\sqrt{k_\perp} \,
	\left(1- \frac{\partial}{\partial \alpha} \right)\,
	I_m(\alpha, \beta, \bm b),
\end{eqnarray}
where the function $I_m(\alpha, \beta, \bm b)$ is given by Eq.~(\ref{eq:I_function}).

Similar to the Yukawa potential, one can employ the amplitude (\ref{eq:amplitude_wave_packet_twisted_hydrogenic_2}) to investigate the scattering of twisted electrons by various hydrogenic targets. Again, we start with a \textit{single} hydrogen atom for which the number of (scattering) events is given by:
\begin{eqnarray}
   \label{eq:number_events_hydrogenic_single}
   \frac{{\rm d}\nu}{{\rm d}\Omega} &=& \frac{N_e a_0^2}{8 \pi \cos\theta_k} \nonumber \\
   && \hspace{-2cm} \times
   \left| \int_0^\infty {\rm d}k_\perp \, g_{\varkappa_0 \sigma_\perp}(k_\perp) \, \sqrt{k_\perp}\,
   \left(1-\frac{\partial}{\partial \alpha}\right)\, I_m(\alpha, \beta,\bm b)
   \right|^2 .
\end{eqnarray}
From this expression and from the properties of the function $I_m(\alpha, \beta, \bm b)$ we can again derive Eqs.~(\ref{eq:angular_distribution_single_Yukawa}) and (\ref{eq:angular_distribution_single_Yukawa_2}) which describe emission pattern of the outgoing electrons for the forward direction and the small impact parameters $b$. Moreover, in the limit of a wide wave--packet, when the momentum distribution function $g_{\varkappa_0 \sigma_\varkappa}(k_\perp)$ is sharply peaked ar $k_\perp = \varkappa_0$, we get from Eqs.~(\ref{eq:eq:numebr_events_single_potential_general_wide}) and (\ref{eq:amplitude_wave_packet_twisted_hydrogenic_2})
\begin{equation}
   \label{eq:number_events_hydrogenic_single_wide}
   \frac{{\rm d}\nu}{{\rm d}\Omega} = L^{({\rm tw})}
   \left| \frac{a_0}{2} \, \left(1 -
   \frac{\partial}{\partial \alpha_0}\right) \, I_m(\alpha_0, \beta_0, \bm b)
   \right|^2 \, .
\end{equation}
Here, $\alpha_0$ and $\beta_0$ are given by Eq.~(\ref{eq:w_alpha_beta_hydrogenic}) with
$k_\perp= \varkappa_0$, and $L^{\rm tw}$  given by Eq. (\ref{eq:Ltw_general}).

With the help of the amplitude (\ref{eq:amplitude_wave_packet_twisted_hydrogenic_2}) one can also treat the scattering of the twisted wave--packet by a macroscopic hydrogenic target. For this case, the angle--differential cross section reads as follows:
\begin{eqnarray}
	\label{eq:number_events_hydrogenic_single_macroscopic}
	\frac{{\rm d}\bar\sigma(\theta,\theta_k,p_f)}{{\rm d}\Omega} &=&
	\frac{a_0^2}{4\cos{\theta_k}}
	\,\int_0^{2\pi} \frac{d\varphi_k}{2\pi} \,
	\Bigg[\frac{1}{u-v\cos{(\varphi_k-\varphi)}} \nonumber \\
	&+& \frac{1}{(u-v\cos{(\varphi_k-\varphi)})^2} \Bigg]^2 \nonumber \\[0.1cm]
	&& \hspace*{-2cm} = \frac{a_0^2}{4\cos{\theta_k}}\left(-\frac{\partial}{\partial u}+
	\frac{\partial^2}{\partial u^2}-\frac{1}{6} \frac{\partial^3}{\partial
		u^3}\, \right)\frac{1}{\sqrt{u^2-v^2}}\,,
\end{eqnarray}
where $u = 1+\mbox{$\frac 12$} a_0^2 p^2_f(1 -\cos{\theta}\cos{\theta_k})$, $v = \mbox{$\frac 12$} a_0^2 p^2_f \sin{\theta}\sin{\theta_k}$, $u > v$, and the wide incident wave--packet is assumed.

%
%
\section{Results and discussions}
\label{sec:results}

In order to illustrate the theory developed in Sections~\ref{sec:theory} and \ref{sec:Yukawa_and_hydrogenic_formulas}, we intend now to present calculations for scattering of the twisted electrons by \textit{hydrogen atoms}. All the results in Figs.~1--4 are presented for the wave packets with the averaged momentum $p_i = 10/a_0$ where $a_0$ is the Bohr radius (this momentum corresponds to the kinetic energy of $1.4$ keV). Similar to before, these calculations are performed for three various scenarios in which electrons collide with a single H atom, as well as with infinitely extended (macroscopic) and localized (mesoscopic) atomic target.

\begin{figure}[t]
	\hspace*{-0.2cm}
	\includegraphics[width=0.99\linewidth]{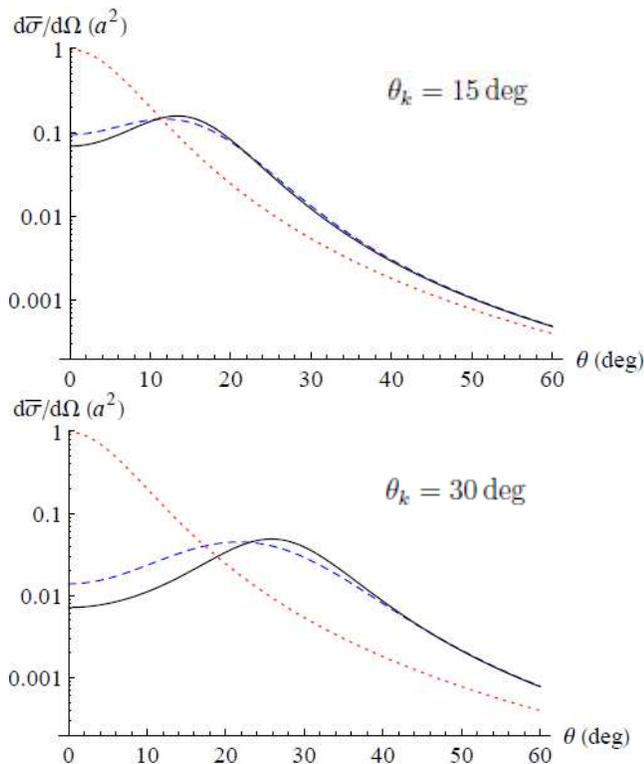}
	\caption{(Color online) The averaged cross section (\ref{eq:cross_section_macroscopic_target}), (\ref{eq:number_events_hydrogenic_single_macroscopic}) for the scattering of twisted electrons by a macroscopic target, consisting of hydrogen atoms in their ground state. Results are presented for the incident wave packets with the width
$\sigma_\varkappa = \varkappa_0/3$ (blue dashed line) and $\sigma_\varkappa \ll \varkappa_0$ (black solid line), and the opening angle $\theta_k =$~15~deg (upper panel) and 30~deg (lower panel). Results of calculations are compared, moreover, with the prediction obtained for the plane--wave electrons (red dotted line).
	\label{Fig1}}
\end{figure}
%
%

%
%
\subsection{Scattering by a macroscopic target}
\label{subsec:results_macroscopic}

We start our calculations from the most experimentally accessible scenario of a macroscopic target. In this case, the transverse extension of an electron beam is assumed to be much smaller than the size of the target, $\mathcal{R} \gg 1/\sigma_{\varkappa}$, and the scattering process is described by the cross section (\ref{eq:cross_section_macroscopic_target}). This cross section, \textit{averaged} over all impact parameters of individual scatterers, is independent of the electron's OAM projection but is still sensitive to the kinematic parameters and to the size of the beam. In order to illustrate such a sensitivity, we display in Fig.~\ref{Fig1} the cross section ${\rm d}{\bar \sigma}/{\rm d}\Omega$ for the scattering of twisted electrons by a (macroscopic) hydrogenic target. Calculations have been performed for two values of the beam opening angle, $\theta_k = 15$~deg (upper panel) and $\theta_k = 30$~deg (lower panel). We also compare the results obtained for the incident plane--wave electrons (red dotted line) with the predictions for the twisted wave--packet (\ref{eq:wavepacket_Bessel_momentum}) with the width $\sigma_{\varkappa} = \varkappa_0/3$ (blue dashed line) and $\sigma_{\varkappa} \ll \varkappa_0$ (black solid line). The latter result corresponds to an approximation of a wide wave--packet which, in its limit, recovers the monochromatic case.

As can be seen in the Fig.~\ref{Fig1}, the angle--differential cross section is indeed very sensitive to the opening angle $\theta_k$. While the incident plane--wave electrons are scattered predominantly in the forward direction, $\theta = 0$~deg, the ${\rm d}{\bar \sigma}/{\rm d}\Omega$ for the twisted wave--packets is peaked near $\theta = \theta_k$. As discussed already in Section~\ref{subsubsec_wide_target} for a wide packet, this behaviour is expected from Eq.~(\ref{eq:cross_section_macroscopic_target_wide_packet_2}) and from the (plane--wave) amplitude $f({\bm Q} - {\bm k}_{\perp}^{(0)})$ that is maximal at $\left({\bm Q} - {\bm k}_{\perp}^{(0)}\right)^2 = 0$. With the increase of $\sigma_\varkappa$ and, hence, decrease of the transversal size of the wave packet $\Delta x = \Delta y \sim 1/\sigma_{\varkappa}$, the ``wide--packet'' approximation is not valid any more. However, even in this ``spatially--localized--packet'' case the twisted electrons are scattered most likely under the angles near $\theta = \theta_k$, although the peak in the emission pattern becomes less sharp, see blue dashed curve in Fig.~\ref{Fig1}.

Despite the different scattering patterns of the incident plane--wave-- and twisted electrons, the \textit{total} cross sections for these two cases are generally of the same order of magnitude. For instance, by integrating the ${\rm d}{\bar \sigma}/{\rm d}\Omega$ (black solid line) and ${\rm d}{\sigma}_{\rm pl}/{\rm d}\Omega$ (red dotted line) from Fig.~\ref{Fig1} over the angle $\theta$, we confirm numerically the relation (\ref{eq:total_cross_section_wide_packet}) between ${\bar \sigma}$ and $\sigma_{\rm pl}$ for the wide twisted packet and the plane wave, respectively. For a spatially localized packet, for which $\sigma_\varkappa \sim \varkappa_0$, the total cross section is about 10--30 \% smaller than the ${\bar \sigma}({\rm wide}) = \sigma_{\rm pl}/\cos\theta_k$, which, again, makes the observation of the scattering of focused twisted beams experimentally feasible.

\begin{figure}[t]
	\includegraphics[width=0.95\linewidth]{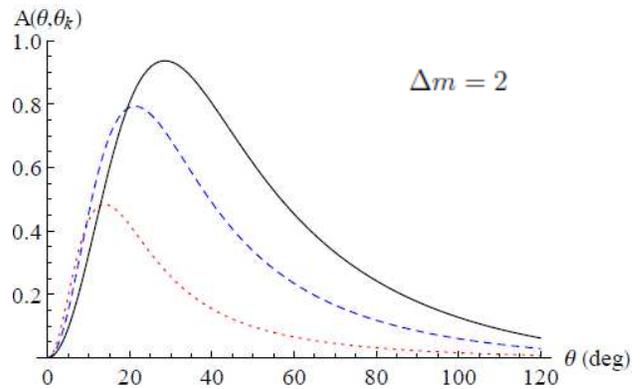}
	\caption{(Color online) The azimuthal assymetry parameter (\ref{eq:cross_section_macroscopic_target_wide_packet_superposition_interference}) for the scattering of the superposition of two Bessel wave--packets by a macroscopic hydrogenic target. The calculations have been performed for two incident electron beams with the width $\sigma_{\varkappa} \ll \varkappa_0$, difference of OAM projections $\Delta m =$~2, and the opening angles $\theta_k =$~10~deg (red dotted line), 20~deg (blue dashed line) and 30~deg (black solid line).
	\label{Fig2}}
\end{figure}

Until now we have discussed the potential scattering of a \textit{single} twisted wave--packet. For this case, and for a macroscopic target, the averaged cross section ${\rm d}{\bar \sigma}/{\rm d}\Omega$ appears to be insensitive to the projection of the orbital angular momentum $m$. However,  the OAM--sensitivity can be restored if the incident electron beam is a coherent superposition (\ref{eq:superposition_two_states}) of \textit{two} twisted states with different $m$'s. As seen from Eq.~(\ref{eq:cross_section_macroscopic_target_wide_packet_superposition_2}), the scattering pattern for such a superposition depends not only on the polar angle $\theta$ but also on the azimuthal one, $\varphi$. This $\varphi$--dependence is parametrized as ${\rm d}{\bar \sigma}^{(2)}/{\rm d}\Omega \sim 1 + A \, \sin\left( \Delta m(\varphi-\pi/2)+\Delta\alpha \right)$ and, hence, is determined by the difference of the beam phases $\Delta \alpha$ and the OAM projections $\Delta m$. The strength of the azimuthal asymmetry is defined by the parameter $A \equiv A(\theta; \theta_k)$ that is given by Eq.~(\ref{eq:cross_section_macroscopic_target_wide_packet_superposition_interference}) and it is insensitive to $\varphi$. In Fig.~\ref{Fig2} we display this parameter $A$ as a function of the polar scattering angle $\theta$ for three opening angles: $\theta_k =$~10~deg (red dotted line), 20~deg (blue dashed line) and 30~deg (black solid line). The calculations have been performed for the \textit{wide} wave packets with $c_1 = c_2 = 1/\sqrt{2}$, $\Delta m = 2$, and the width $\sigma_{\varkappa} \ll \varkappa_0$. As seen in the figure, the $A(\theta; \theta_k)$ is peaked at the angles $\theta \approx \theta_k$ and, moreover, it increases with the growth of $\theta_k$. This implies that 
the $\Delta m$-- and $\Delta \alpha$--dependences can be observed most easily for the beams with large opening angles $\theta_k$ and for electron detectors placed at $\theta \approx \theta_k$.

\begin{figure}[t]
	\includegraphics[width=0.95\linewidth]{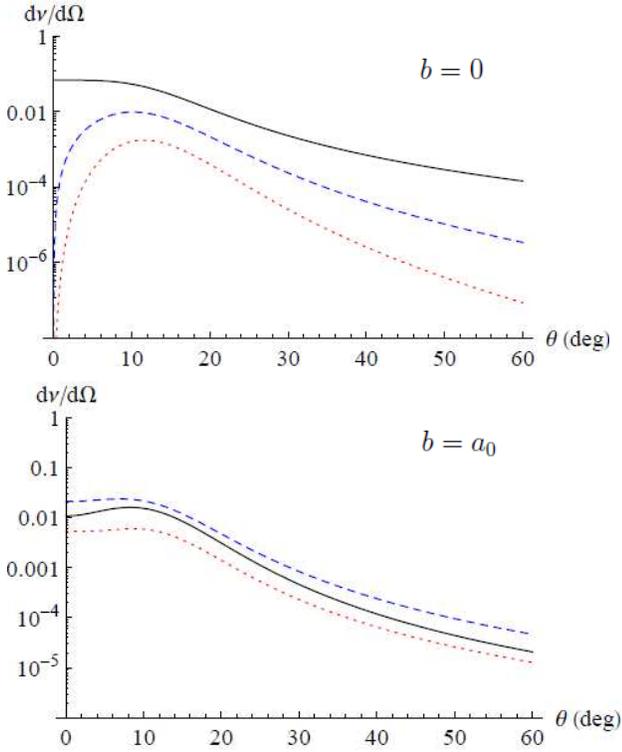}
	\caption{(Color online)
		The (normalized) number of events ${\rm d}\nu(\theta) / {\rm d} \Omega $ with $N_e=1$ from Eq.~(\ref{eq:number_events_hydrogenic_single})
for the scattering of the Bessel wave--packet by a single hydrogen atom, placed at a distance $b = 0$ (upper panel) and $b = a_0$ (lower panel). Results are presented for the incident packets with the width $\sigma_\varkappa =  \varkappa_0/5$, the opening angle $\theta_k =$~10~deg, and the OAM projection $m = 0$ (black solid line), $m = 1$ (blue dashed line), and $m = 2$ (red dotted line).
		\label{Fig3}}
\end{figure}
%
%

%
%
\subsection{Scattering by a single potential}
\label{subsec:results_single}

Until now we have discussed the most realistic (experimental) scenario is which twisted electron wave--packet collides with a macroscopic atomic target. However, in order to better understand the features of the potential scattering of Bessel packets it is more convenient to consider a collision with a \textit{single} atom, located at a particular impact parameter $b$ with respect to the beam's axis. As mentioned above, the number of (scattering) events ${\rm d}\nu / {\rm d} \Omega$ can be used to characterize the process for such a ``single--atom--case''. In Fig.~\ref{Fig3} we display, for example, the (normalized) number of events ${\rm d}\nu(\theta) / {\rm d} \Omega $ with $N_e=1$  from Eq.~(\ref{eq:number_events_hydrogenic_single}), as a function of the polar angle $\theta$ of outgoing electrons, defined with respect to the beam ($z$--) axis. Here we assume $\varphi = 0$, that is, scattered electrons are ``detected'' within the plane of the target atom. Calculations were performed for a hydrogen in the ground $1s$ state and for the incident wave packet with the width $\sigma_\varkappa = \varkappa_0/5$. We consider, moreover, two impact parameters, $b = 0$ (upper panel) and $b = a_0$ (lower panel) and three projections of the orbital angular momentum, $m = 0$ (black solid line), $m = 1$ (blue dashed line), and $m = 2$ (red dotted line).

As seen from the Fig.~\ref{Fig3}, there is a strong dependence of the electron scattering pattern on the OAM projection. This $m$--dependence is most pronounced for the central collision, $b = 0$, and small angles, $\theta \to 0$. In this case, the (forward) electron scattering is allowed only for $m = 0$, while ${\rm d}\nu (\theta=0)/{\rm d} \Omega$ vanishes identically for $m = 1$ and $m = 2$. Such a behaviour is expected from the analysis of the transverse component of the electron wave--function and of the scattering amplitude, see Eq.~(\ref{eq:amplitude_wave_packet_b_0_forward}). With the increase of the impact parameter $b$ the dip in the electron angular distribution for $\theta = 0$ (and $m \ne 0$) disappears. Indeed, for $b = a_0$ (lower panel of Fig.~\ref{Fig3}) the forward scattering is allowed and even becomes dominant as the target atom is further shifted from the center of the incident packet.

In contrast to the case of the macroscopic target, the collision of a wave--packet having one definite value of $m$ with a single well--localized atom can also result in the \textit{azimuthal asymmetry} of the electron scattering pattern.
\begin{figure}[t]
	\includegraphics[width=0.9\linewidth]{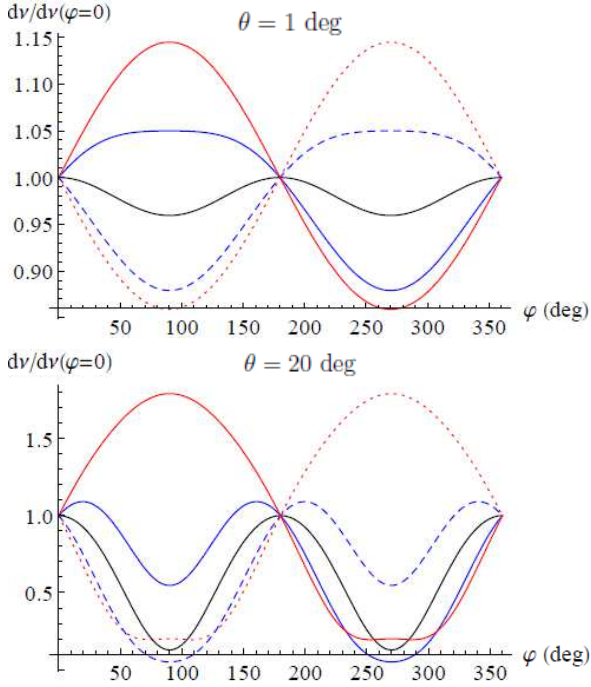}
	\caption{(Color online)
		The azimuthal angle dependence of the normalized number of events ${\rm d}\nu (\theta, \varphi)/ {\rm d}\nu (\theta, \varphi = 0)$ for scattering of the Bessel wave--packet by a single hydrogen atom, placed at the distance $b = 2a_0$. The results are presented for the incident packets with the width $\sigma_\varkappa = \varkappa_0/5$, the opening angle $\theta_k =10$~deg, and the OAM projection $m = -2$ (red dotted line), $m = -1$ (blue dashed line), $m = 0$ (black solid line), $m = 1$ (blue solid line), and $m = 2$ (red solid line). We assumed, moreover, that outgoing photons are detected at the polar angles $\theta = 1$~deg (upper panel) and 20~deg (lower panel).
		\label{Fig4}}
\end{figure}
One can expect this because the system of the incident packet \textit{plus} target atom at $b \ne 0$ does not possess the azimuthal symmetry, that is ``recovered'' only after integration over $b$. The resulting $\varphi$--distribution of outgoing electrons
(see Eq.~(\ref{eq:number_events_hydrogenic_single})) is very sensitive to the kinematic parameters and the OAM projection of Bessel electrons. In Fig.~\ref{Fig4}, for example, we display the normalized number of scattered events, ${\rm d}\nu (\theta, \varphi)/ {\rm d}\nu (\theta, \varphi=0)$, for the incident wave--packet whose parameters  are $\theta_k = 10$~deg, and $\sigma_\varkappa = \varkappa_0/5$. Here, we performed calculations for the impact parameter $b = 2 a_0$, two polar scattering angles, $\theta = 1$~deg (upper panel) and $\theta = 20$~deg (lower panel), and five OAM projections:
$m = -2$ (red dotted line), $m = -1$ (blue dashed line), $m = 0$ (black solid line), $m = 1$ (blue solid line) and $m = 2$ (red solid line).
As seen from the figure, the variation of $m$ may lead to the \textit{qualitative} changes in the azimuthal angular distribution, thus suggesting that the scattering by well--localized targets can be used for diagnostics of the twisted beams.

%
%
\subsection{Scattering by a mesoscopic target}
\label{subsec:results_mesoscopic}

In Sections~\ref{subsec:results_macroscopic} and \ref{subsec:results_single} above we treated two limiting cases of scattering either by an infinitely large target or by a single atom. These calculations have clearly indicated that the size of a target strongly influences the OAM--sensitivity of the electron scattering pattern. Indeed, while the averaged cross section (\ref{eq:cross_section_macroscopic_target}) is independent of $m$ unless the superposition of two packets is considered, the number of events (\ref{eq:numebr_events_single_potential_general}) for a single potential varies significantly with the change of the OAM. In order to illustrate better the ``size--effect'' in scattering of the twisted wave--packets, here we apply Eqs.~(\ref{eq:number_of_events_localized_target}) and (\ref{eq:Gaussian_distribution_2}) which describe collisions with a target of finite sizes. By making use of these expressions, we have evaluated the number of scattering events for
two limiting cases given in Sect. II C 4.
\begin{figure}[t]
	\includegraphics[width=0.9\linewidth]{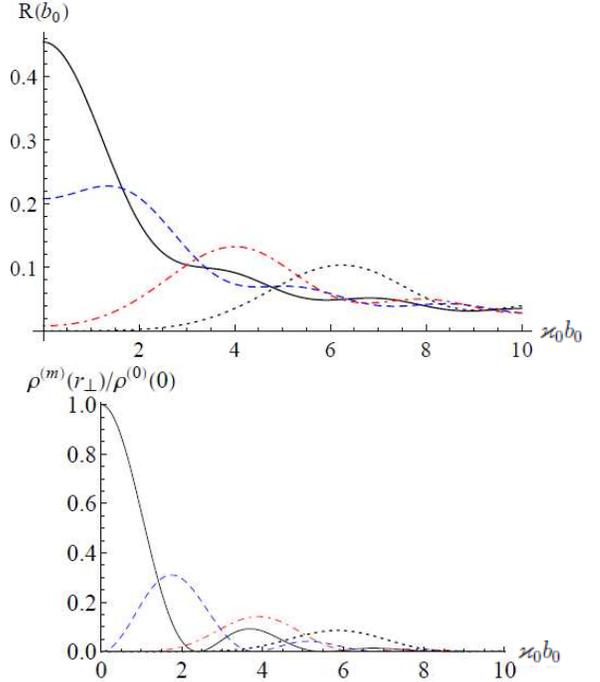}
	\caption{(Color online)
 The relative differential cross section
 $R(b_0)=d\sigma^{\rm (mesos)}(b_0)/d\sigma^{\rm (PW)}$ from Eq.~(\ref{ratioRb0}) as a function of the impact parameter $b_0$ of the target center (upper panel) for $\sigma_b=1/\varkappa_0=10$ nm, $1/\sigma_\varkappa = 50$ nm, $\theta_k\ll 1$ and for the following projections of the orbital angular momentum: $m=0$ (black solid line), $m=1$ (blue dashed line), $m=3$ (red dot-dashed line), $m=5$ (black dotted line). On the lower panel the normalized density of the incident twisted beam $\rho^{(m)}(r_\perp)/\rho^{(0)}(0)$  is shown for the same parameters.
		\label{Fig5}}
\end{figure}

When the size of the target $\sigma_b$ is smaller than  the transverse spread of the wave--packet $\Delta x = \Delta y \sim 1/\sigma_{\varkappa}$, we use Eq.~(\ref{ratioRb0}) and present in Fig.~\ref{Fig5} (upper panel) the relative differential cross section
 \be
 R(b_0)=\fr{{d\sigma^{\rm (mesos)}(b_0)}/{d\Omega}}
 {{d\sigma^{\rm (PW)}}/{d\Omega}}
  \ee
as a function of the impact parameter $b_0$ of the target 
center. Calculations have been performed for the following parameters: $\sigma_b=1/\varkappa_0=10$ nm, $1/\sigma_\varkappa = 50$ nm, $\theta_k\ll 1$ and for four projections of the orbital angular momentum: $m=0$ (black solid line), $m=1$ (blue dashed line), $m=3$ (red dot-dashed line), $m=5$ (black dotted line). On the lower panel of Fig~\ref{Fig5} we present the normalized density of the incident twisted beam $\rho^{(m)}(r_\perp)/\rho^{(0)}(0)$ for the same parameters. Here the density itself reads as follows (see Eq.~(\ref{eq:transverse_component_convolution})):
\be
 \rho^{(m)}(r_\perp)= \left|\int_0^\infty \sqrt{\varkappa/(2\pi)}\,
 J_m (\varkappa r_\perp) g_{\varkappa_0 \sigma_\varkappa}\varkappa\,
 {\rm d} \varkappa \right|^2.
\ee
It is clearly seen from comparison of these panels that the ``twisted'' cross section is very sensitive to variation of the incident beam's density.

When the size of the target $\sigma_b$ is larger than the transverse spread of the wave--packet $1/\sigma_{\varkappa}$, we use Eq.~(\ref{eq:large-beam}) and
present in Fig.~ 6 the number of scattering events as a function of the impact parameter for three different values of the OAM: $m=0$ (black solid line), $m=50$ (blue dashed line), $m=100$ (red dotted line). Clearly, scattering off the large target is sensitive to the spatial density of the incident wave front only for the big values of the OAM when $\sigma_b \sim m/\varkappa_0,\, m \gg 1$.
Note that electrons with the OAM quanta as high as $m=200$ have already been generated \cite{GrG15}.
\begin{figure}[t]
	\includegraphics[width=0.99\linewidth]{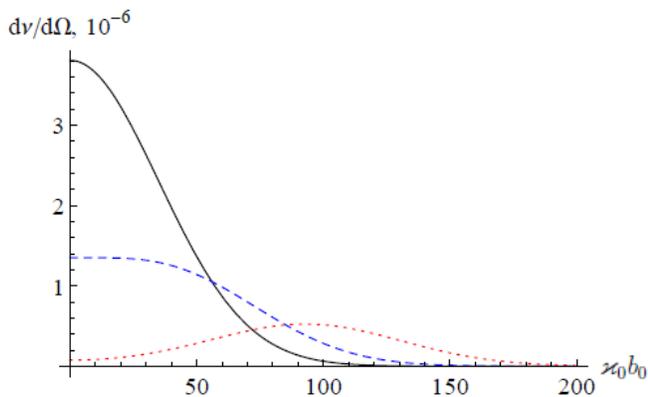}
	\caption{(Color online)
 The number of events (\ref{eq:large-beam}) for a wide target as a function of the impact parameter $b_0$ for $\sigma_b = 10$ nm, $1/\sigma_\varkappa = 2$ nm, $\theta_k = \theta = 1\, \text{deg},\, \varphi = \varphi_b = 0$ and for the following projections of the OAM: $m=0$ (black solid line), $m=50$ (blue dashed line), $m=100$ (red dotted line).
		\label{Fig6}}
\end{figure}
%
%
%
%
\section{Summary}
\label{sec:summary}

In summary, we have applied the generalized Born approximation, developed by us in Ref.~\cite{KaK16}, in order to investigate scattering of the vortex electrons by atomic targets. In our study, we focused especially on derivation of the physically meaningful expressions for the number of (scattering) events and for the differential cross sections. These physical observables have been obtained for different ``experimental'' scenarios in which incident wave--packet collides with (i) a single potential as well as with (ii) localized (mesoscopic) target and with (iii) an infinitely wide (macroscopic) one. Even though the developed theory can be employed for \textit{any} type of the electron--atom interaction, in the present study we described the scattering off the Yukawa- and hydrogenic potentials. For these potentials, simple expressions for the cross sections and for the numbers of scattering events are presented which can be used for analysis and guidance of the future scattering experiments.

On the basis of the developed theory we showed that the number of scattering events in collisions involving twisted electrons is comparable to that in the standard plane--wave regime. This implies that experiments with the focused twisted electrons are feasible with the present-day detectors. The outcome of these experiments will depend, however, on the relative size of an incident beam and an atomic target. For example, for targets whose width is narrower than the transverse size of the beam, the angular distribution of the scattered electron can be very sensitive not only to the opening angle, but also to the OAM projection itself. The sensitivity to the orbital momentum $m$ vanishes, however, with the increase of the target's size. 
On the other hand, even in this macroscopic case the OAM--sensitivity can be recovered if one \textit{prepares} an incident beam as a coherent superposition of two twisted packets with two different $m$'s. We conclude, therefore, that the potential scattering can provide a wide range of opportunities for diagnostics of the twisted electron beams, and the corresponding experiments can be carried out at the present--day facilities.

\

D.K. wishes to thank C.\,H.\,Keitel, A.\,Di Piazza and S.\,Babacan for their hospitality
during his stay at the Max-Planck-Institute for Nuclear Physics in Heidelberg.
D.K. is supported by the Alexander von Humboldt Foundation (Germany)
and by the Competitiveness Improvement Program of the Tomsk State University. G.L.K and V.G.S. are supported by the Russian Foundation for Basic Research via grant 15-02-05868.

%
%
%
%

\end{document}